# High-quality hexagonal boron nitride selectively grown on patterned epigraphene by MOVPE

*Vishnu Ottapilakkal[1,#], Abhishek Juyal[1,2,#], Suresh Sundaram[1,3,4], Phuong Vuong[1,3], Collin Beck[2], Noel L. Dudeck[2], Amira Bencherif[1,2], Annick Loiseau[5], Frédéric Fossard[5], Jean-Sebastien Mérot[5], David Chapron[6], Thomas H. Kauffmann[6], Jean-Paul Salvestrini[1,3,4], Paul L. Voss[1,4], Walt A. de Heer[2], Claire Berger[1,2,\*], Abdallah Ougazzaden[1,4\*]*

[1]CNRS, IRL 2958 Georgia Tech - CNRS, 2 rue Marconi, 57070 Metz, France

[2]Georgia Institute of Technology, School of Physics, Atlanta, GA 30332-0250, USA

[3]Georgia Tech-Europe, 2 rue Marconi, 57070 Metz, France

[4]Georgia Institute of Technology, School of Electrical and Computer Engineering, Atlanta, GA 30332-0250, USA

[5]Laboratoire d'Etude des Microstructures, ONERA-CNRS, Université Paris Saclay, F-92322 Châtillon, France

[6]Laboratoire Matériaux Optiques, Photonique et Systèmes (LMOPS), Université de Lorraine & Centrale Supélec, 57070 Metz, France

ABSTRACT

Hexagonal boron nitride encapsulation is the method of choice for protecting graphene from environmental doping and impurity scattering. It was previously demonstrated that metal-organic vapor phase epitaxy (MOVPE) grows epitaxially ordered, uniform BN layers on epigraphene (graphene grown on SiC). Due to graphene's non-wetting properties, h-BN growth starts preferentially from the graphene ledges. We use this fact here to selectively promote growth of high-quality flat h-BN on epigraphene by patterning epigraphene microstructures prior to BN growth. Thin h-BN films (down to 6 nm) grown by MOVPE show smooth and pleated surface morphology on epigraphene, while crumpled BN is observed on the SiC. Cross-sectional high-resolution transmission electron microscopy images and fluorescence imaging confirm the higher BN quality grown on the epigraphene. Transport measurements reveal p-doping as expected from hydrogen intercalation of epigraphene and regions of high and low mobility. This method can be used to produce structurally uniform high-quality h-BN/epigraphene micro/nano scale heterostructure.

INTRODUCTION

Two-dimensional (2D) materials such as graphene, hexagonal boron nitride (h-BN), and transition metal dichalcogenides (TMDCs) have attracted significant interest in recent years due to their unique physical properties and potential for electronic and photonic applications.[1,2] Epigraphene (EG) is structurally high quality graphene grown epitaxially on SiC[3–6], with exceptional properties such as quantized ballistic electronic transport on length scales of hundreds of micrometers[7], highly efficient spin transport[8], high frequency transistor operation[9], quantum Hall effect-based standard of resistance[10] and it is compatible with largescale nanoelectronics processing methods.[3] Ultra-high mobility semiconducting epigraphene was recently demonstrated in the epigraphene buffer layer, with mobilities up to 5000 cm$^2$/Vs [11], which is 20 times the theoretical limit of other 2D semiconductors. The required single crystal SiC substrates are commercially available in up to 200 mm wafers, as they are widely used for high temperature /high power electronics and compatible with THz electronics applications.[12]

To fully exploit epigraphene potential properties, it requires to be integrated with dielectrics, while sustaining graphene pristine properties in an industrially relevant process. Exfoliated h-BN is a widely-used dielectric capping layer to protect graphene from environmental doping and to mitigate impurity scattering[13–15]. h-BN is a 6 eV wide bandgap semiconductor with a dielectric constant of about 4.[16] It is isostructural with graphene (1.7% lattice mismatch), charge neutral, doesn't form chemical bonds with graphene. It has high temperature and chemical stability. We have previously demonstrated a process where Migration Enhanced Metal Organic Vapor Phase Epitaxy (ME-MOVPE) of h-BN on epigraphene by van der Waals growth produces epitaxially ordered h-BN/EG/SiC heterostructures of excellent quality, with flat and clean interfaces.[15] This process is scalable and compatible with industrial methods so that it has significant advantages over the conventional mechanical transfer processes.

The ME-MOVPE BN growth process has been explained by a lateral epitaxial deposition mechanism involving a one-dimensional nucleation-free-energy-barrierless growth that relies on the large mobilities of the B and N species on the graphene surface.[15] The method was inspired by our observation that on partially grown epigraphene on SiC, flat h-BN layers grow from the graphene edge, whereas a large density of crumpled BN particles are observed on (oxidized) SiC and also accumulate on defects and step edges.[15] Here we use this fact to promote growth of flat h-BN selectively on graphene by patterning epigraphene microstructures prior to h-BN deposition. The process was specifically developed to coat micro/nano scale devices (e.g. Hall-bars) with h-BN. Their small lateral size steers high-quality h-BN growth, as growth of 2D materials on micrometre-sized patterned substrates results in homogeneous nucleation, leading to improved crystal quality.[17] The small size will also make the underlying epigraphene more likely to be defects- and step- free.

EXPERIMENTAL METHODS

Epigraphene was produced on commercial chemical mechanical polished (CMP) insulating 4H-SiC substrates (from II-VI, now Coherent) by the polymer enhanced confinement-controlled sublimation growth method.[18] The growth temperature is adjusted around 1600°C under 1 atm Ar. Details of the epigraphene growth have been reported elsewhere.[18–20] Epigraphene microstructures were fabricated by standard lithography: a 300 nm thick ma-N-2403 resist was patterned via electron beam lithography and epigraphene was etched with oxygen plasma (RIE) at a pressure of 15 mTorr, an oxygen flow 4 sccm and an RIE power of 16 W for 15s.

BN growth was carried out using an Aixtron MOVPE close-coupled showerhead (CCS) 3x2" reactor. Triethylboron and ammonia were used as the precursors for boron and nitrogen, respectively. The growth was carried out at 1270°C in a hydrogen environment at 85 mbar, following the similar growth conditions as previously reported.[15,21,22] Raman spectroscopy measurements and fluorescence emission mapping were performed using Horiba Lab RAM HR equipment with an excitation wavelength of 532 nm. The fluorescence emission mapping was acquired with a lateral resolution of 1 μm by exciting the sample with a 532 nm laser and measuring the fluorescence emission spectra in the range of 570 nm to 630 nm. High-resolution scanning transmission electron microscopy (HR-STEM) was performed using a probe-corrected ThermoFisher Scientific Titan G2 on cross-sections prepared by focused ion beam (FIB). Atomic force microscopy (AFM – Bruker Edge) in non-contact mode and scanning electron microscopy (SEM – Zeiss Supra 55VP with an SE2 detector) was utilized for morphological characterization.

RESULTS AND DISCUSSION

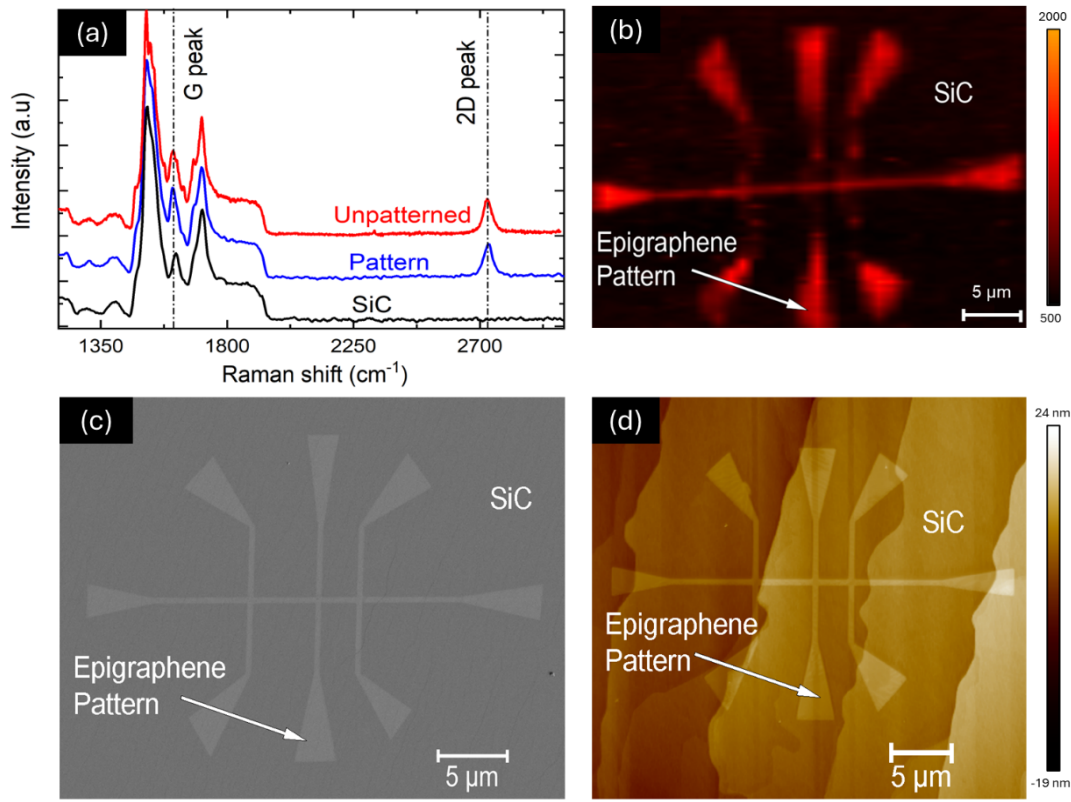

**Figure 1.** (a) comparison of Raman spectra before (red) and after (blue) patterning, highlighting the unchanged characteristic epigraphene peaks after patterning. A bare SiC spectrum (black) is shown for comparison, (b) Raman mapping (2D peak) of the sample revealing the Hall bar-shaped epigraphene structure, (c) SEM and (d) AFM images of epigraphene Hall bar. The SiC terraces are visible onto which the epigraphene grows continuously.

The structural characteristics and morphology of patterned epigraphene are presented in Figure 1. The Raman spectrum remains the same before (red trace) and after (blue trace) patterning process (Figure 1a), indicating that the quality of epigraphene is unaltered by the patterning process. The characteristic 2D-peak [23] of graphene at 2738 cm$^{-1}$ show a compressive strain as is usual for epigraphene on the Si-face[24,25], and a FWHM less than 40 cm$^{-1}$ indicates monolayer graphene.[23,24] The graphene G peak at 1600 cm$^{-1}$ is well resolved, despite the large SiC peaks in that energy range. The absence of a Raman D-peak around 1350 cm$^{-1}$ on both spectra confirm the high quality of the epigraphene. A bare SiC Raman spectrum (black trace) is included for comparison. Raman mapping of the 2D peak, shown in Figure 1b, confirms the successful removal of epigraphene outside of the Hall bar. The SEM and AFM images in Figure 1c, d respectively, show a well-defined patterned Hall-bar structure, with the continuity of the epigraphene clearly visible in the AFM image despite the presence of SiC steps.[24] The AFM roughness of the epigraphene throughout the structure after patterning is less than 0.2 nm.

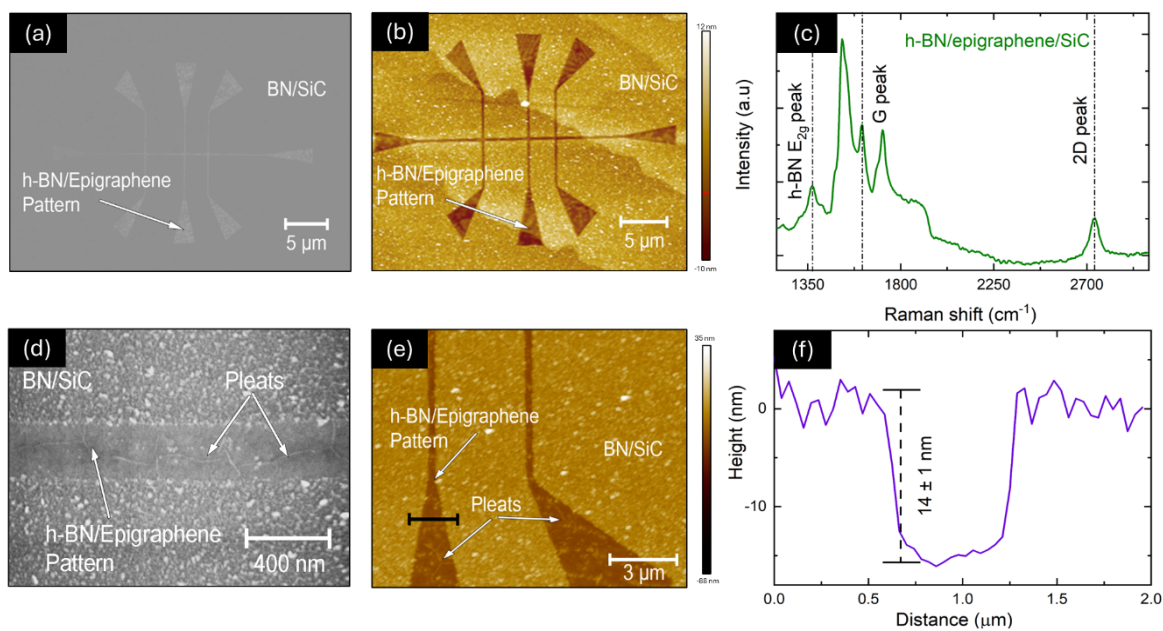

**Figure 2.** (a) SEM and (b) AFM image of the BN grown on patterned epigraphene sample. (c) Raman spectrum recorded on the heterostructure pattern, clearly indicating the presence of both the epigraphene and h-BN, high-magnification (d) SEM and (e) AFM

image of the heterostructure showing the pleated h-BN on the patterned epigraphene and granular BN on SiC substrate. (f) Height profile measured along the black line in (e), showing a dip in the BN surface across the pattern.

The patterned epigraphene samples were transferred to the MOVPE reactor for h-BN growth. The growth was adjusted to grow 6-30 nm thick h-BN, by tuning the deposition time. The Scanning Electron Microscopy (SEM) and Atomic Force Microscopy (AFM) images in Figure 2a, b respectively, reveal that after BN growth, on the epigraphene patterned structure, a very smooth surface was obtained, which is different from the surface deposits outside of the pattern (i.e. on $O_2$ plasma etched SiC). The flat surface on the epigraphene structure is contrasted with granular and rough surface on the (oxidized) SiC, as previously observed in ref [15].

We first confirm the presence of BN on the epigraphene patterns with Raman spectroscopy. In addition to the graphene (2D peak at 2734 cm$^{-1}$ and G peak at 1600 cm$^{-1}$) and SiC peaks, a prominent peak is observed at 1370 cm$^{-1}$ at the expected location of the h-BN $E_{2g}$ peak (as shown in Figure 2c). This peak is close to the position of the graphene D peak (1350 cm$^{-1}$), superimposed with SiC features, making it difficult to resolve. However, the D peak is not detectable in as-grown epigraphene samples (Figure 1a), and no swelled D-peak was observed in the underlying graphene after peeling off the BN film.[15] Therefore, we confidently ascribe the peak observed at 1370 cm$^{-1}$ to h-BN $E_{2g}$. In addition, considerable background fluorescence was observed in the Raman spectrum in the range 1200-2000 cm$^{-1}$ after BN growth, which is not seen in the Raman spectra of bare epigraphene shown in Figure 1a, this also indicates the presence of h-BN and is probably due to point defects in the h-BN layers.[26,27]

The high-magnification SEM and AFM images of Figure 2d, e show pleats on the h-BN on the epigraphene surface. These pleats are characteristic of a thin 2D film grown at high temperature on a substrate that contracts differently upon cooling, as expected for 2D h-BN layers grown at 1270°C on epigraphene/SiC. Note that monolayer epigraphene on the Si-face doesn't show pleats (strained graphene as indicated by the Raman 2D peak blue-shift) and graphene doesn't grow on SiC under the MOVPE h-BN deposition conditions.[15] The surface roughness, as determined from the AFM image of h-BN on patterned epigraphene, averages approximately 0.7 nm, while the pleat heights range between 4 and 8 nm.

In contrast, the granular BN on SiC has a measured AFM roughness of about 3 nm. The granular morphology and the higher AFM roughness could be due to a higher nucleation density of BN because of the lower adatom mobility, which is consistent with the expected effects of the rough surface morphology of (oxidized) SiC.[28] The dip in the AFM height profile across the h-BN/epigraphene pattern (Figure 2f) is expected from the crumpled BN layers on the SiC taking up more space than flat layers on the epigraphene. Also bare SiC outside of the epigraphene swells upon oxidation during the $O_2$ plasma process used to remove graphene and define the Hall bar. We cannot exclude a difference in growth rate of BN on epigraphene and SiC under these growth conditions that can be related to the change in surface free energy of the substrates/templates.[29,30]

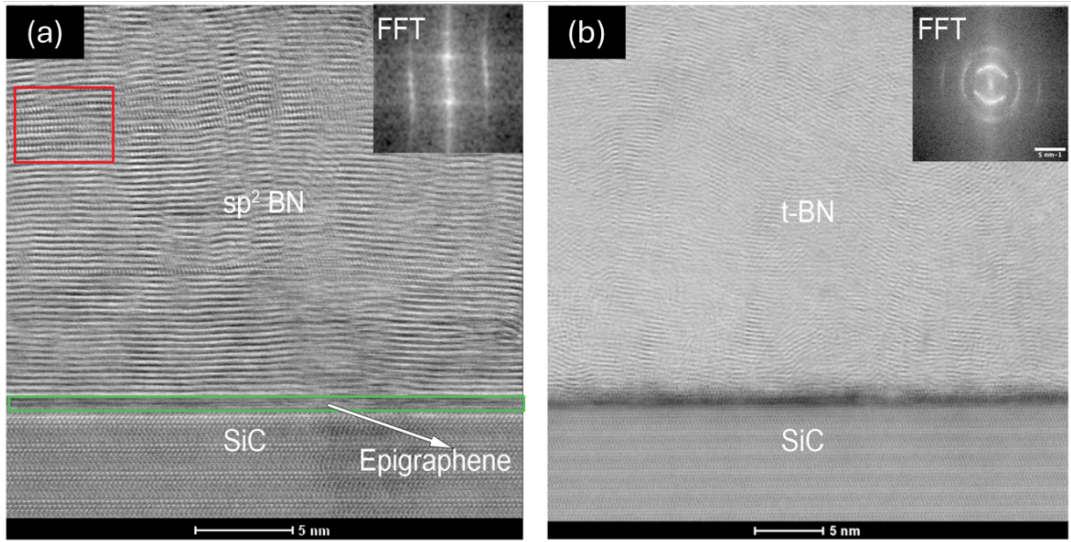

**Figure 3.** Cross-sectional TEM images of (a) h-BN/epigraphene/SiC heterostructure showing the BN layer on 2-layer graphene (buffer + graphene) and inset is the Fast Fourier transform (FFT) of the region enclosed in the red box, and (b) BN/SiC depicting the randomly oriented BN layers grown on SiC substrate (inset: FFT pattern of the region enclosed in the red box indicating the random arrangement of layers).

HR-STEM images on cross-sections of 20 nm h-BN grown on patterned epigraphene prepared by Focused Ion Beam (FIB) are shown in Figure 3. The studies were conducted on both BN/epigraphene patterns on SiC and BN/SiC, as depicted in Figure 3a, b, respectively. The BN grown on the epigraphene patterns exhibits a layered structure consisting of a stacking of almost flat layers with interatomic spacing of d = 0.34 nm, which is consistent with interatomic distances reported in the literature.[22] The spotty FFT pattern clearly shows that the layers have a long-range order. Although BN layers exhibit an orderly stacking arrangement, determining the stacking sequence from the FFT pattern (shown in the inset of Figure 3a) was difficult. In contrast, the BN layers grown outside of graphene (i.e. SiC) showed turbostratic stacking as clearly observed in the HR-STEM image (Figure 3b), and from the arcs of diffuse scattering in the FFT pattern (inset of Figure 3b).

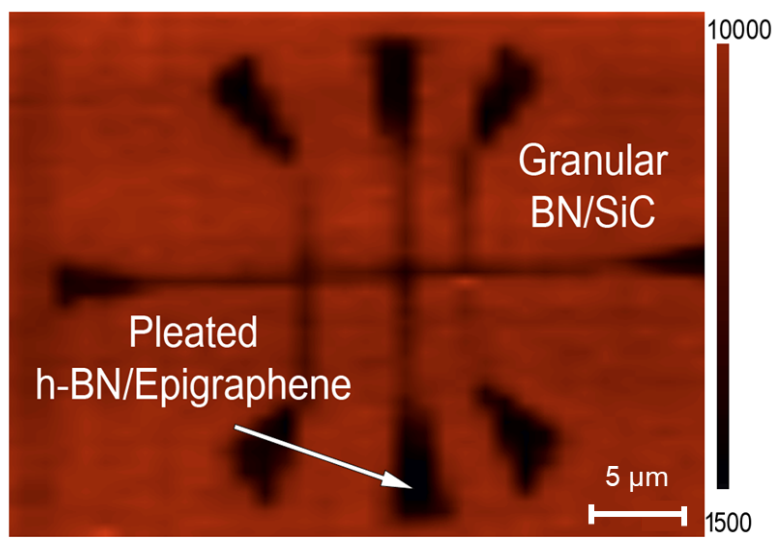

**Figure 4.** Fluorescence emission mapping performed on the heterostructure showing the different intensity from h-BN/epigraphene and BN/SiC.

Fluorescence emission mapping was used to study the difference in quality between BN grown on epigraphene and on SiC. The fluorescence emission intensity was measured range (570–630 nm). Following Refs. [21,31,32] we attribute the fluorescence emission in this range to defects (mostly due to the carbon incorporation[32]) in the BN layer. Figure 4 shows fluorescence emission map of a nominal 6 nm thick BN on patterned epigraphene on SiC sample. A very low fluorescence intensity (almost 7 times) was observed on the h-BN/epigraphene structure compared to the BN/SiC. This low intensity fluorescence on h-BN on epigraphene confirms that h-BN film grown on the epigraphene exhibits better quality and is possibly thinner than on SiC. The selective growth of high-quality h-BN on epigraphene and crumpled BN on (oxidized) SiC is similar to BN grown on sapphire substrates coated with a patterned $SiO_2$ layer, where high-quality h-BN grows on the bare sapphire and BN crumpled particles are observed on the $SiO_2$-covered areas.[33] Similar results are observed on other substrates; for instance, different orientations of Ni substrates also results in BN epilayer of different quality.[29,30]

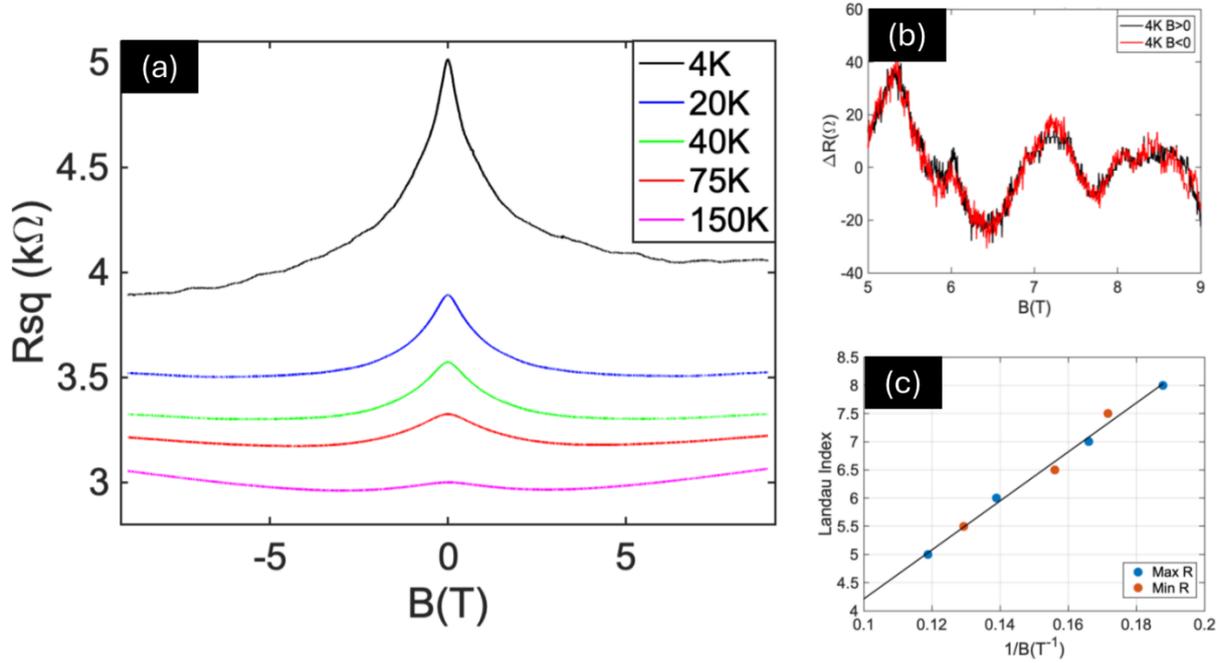

**Figure 5.** Sample 1: 30 nm BN deposited on a graphene patterned Hall bar 6 x15μm. (a) square magnetoresistance at various temperatures showing a large zero field peak that decreases with temperature. (b) Shubnikov de Haas oscillations outlined by subtracting a smooth background from the data in (a). The traces for positive and negative field exactly superimpose at high field. (c) Fan plot of the field of the maxima and minima in (b). The index values are shifted by ½ for the minima.

Transport measurements were performed on several samples patterned with Hall bar devices similar to that of Figure 2 and on a non-patterned sample. For sample 1, a 6 μm x 15 μm Hall bar, we find a large resistivity $R_{sq} = 4k\Omega$, a positive charge density $n_s = 6 \times 10^{12}$cm$^{-2}$, determined from the linear in field Hall effect (Figure S1) and a relatively low mobility value $\mu \sim 250$ cm²/Vs. The magnetoresistance (Figure 5a) presents a broad peak at zero field that decreases with temperature, suggestive of weak localization (WL). According to WL theory, the peak width of about 3 T indicates a phase coherence $l_\phi$ of about 30 nm[34,35], larger, as required, than the scattering length calculated from the resistivity ($l_\varepsilon \sim 5$ nm). Similar resistances and charge densities are measured by the Van der Pauw method[36] on a non-patterned macroscopic sample (Sample 2: 3 x 4 mm², see Figure S2), indicating that the low mobilities are not related to the graphene patterning process.

At low temperature and for B larger than 4 T, small oscillations are superimposed to the broad peak, with maxima (minima) at the same magnetic field value for both field directions. Small deviation from linearity in field is also observed in the Hall effect at 4 K (note that, as expected at such high charge density, no quantum Hall effect is observed, Figure S1). Figure 5c is a plot of the Landau level index N as a function of the values of $1/B_N$ corresponding to the resistance maxima after subtraction of a smooth background (Figure 5b): $N + \gamma = \frac{B_0}{B_N}$. The $B_N$ values for the resistance minima were also included on the figure, by shifting N by ½. The linearity of the plot is a strong indication that these are Shubnikov de Haas oscillations. The charge density can be calculated from the slope $B_0 = n_s \frac{h}{4e}$. We find $n_s = 4.2 \times 10^{12}$cm$^{-2}$, consistent with the charge density determined from the Hall effect (Figure S1). The lowest index level (N=5 at 8.5 Tesla) is also consistent with a projected $R_{xy} = (h/4e^2)(N + 1/2)^{-1} \approx 1k\Omega$, close to the measured value at that field. The $\gamma = 0.1$ value, found by extrapolation to $1/B_N=0$, is compatible with single layer graphene[4,37,38] but this estimate is not very reliable, given the small number of points and the relatively broad maxima. The important point here is that Shubnikov de Haas oscillations can only been observed for $\mu B \gtrsim 1$[39]. This indicates a mobility $\mu > 2,500$ cm$^2$/Vs, inconsistent with the low mobility values found from the resistance. We therefore conclude that the samples present inhomogeneities, with regions of high and low mobility. Similar conclusion was reached for free standing bilayer epigraphene, where the SiC substrate was etched away under the graphene.[40]

Positive charge carrier could indicate hydrogen intercalated epigraphene (a.k.a. quasi-free-standing graphene).[41,42] This is expected here from the BN deposition process, since by annealing above 900 ºC in a hydrogen containing environment, hydrogen atoms passivate the SiC surface, decoupling the first graphene layer (buffer layer) at the SiC interface. This results in a larger SiC-first graphene layer distance, as was indicated in previously measured HR-STEM cross sectional image of similarly prepared BN/EG/SiC heterostructures.[15]

Further work will be necessary to understand the origin of electronic scattering. Cross-sectional TEM indicates that the BN/epigraphene interface is epitaxial and clean. The high temperature deposition process in a hydrogen carrier gas is beneficial to clean the graphene surface from impurities deposited on the graphene during transfer from the growth furnace to the deposition chamber. On the other, we cannot exclude damage to the epigraphene by the hydrogen carrier gas, despite the absence of a noticeable Raman D-peak for the graphene under BN.[15] Inhomogeneous hydrogen intercalation is possible. Another possibility would be scattering induced by defects in the as-deposited BN. The next step will be measuring a high temperature annealed BN/patterned epigraphene/SiC heterostructure as we know this results in higher quality BN layers (sharper BN X-ray peak and no fluorescence in Raman spectroscopy).[21]

CONCLUSION

In summary, we have demonstrated the selective growth of smooth and continuous high-quality h-BN on micrometer-sized monolayer epigraphene patterns by MOVPE method down to a thickness of 6 nm. In between the h-BN/epigraphene heterostructures, the BN layers are crumpled with granular morphology on the SiC substrate, denoting a high nucleation rate. In particular, fluorescence mapping shows the higher quality and possibly smaller thickness of h-BN layers grown on epigraphene than on SiC. Transport results indicate inhomogeneities with low and high mobility regions and a high positive charge density. The presence of high mobility areas in particular is extremely promising and further improvement can be expected by high temperature annealing, which further improves BN quality, and by depositing BN on step free graphene on atomically flat terraces that can now be produced on mm scale.[11] These results show that MOVPE is a promising technique for growing structurally uniform h-BN for graphene-based devices.

ASSOCIATED CONTENT

Supporting Information

AUTHOR INFORMATION

**Corresponding Author**

*Author to whom correspondence should be addressed: abdallah.ougazzaden@georgiatech-metz.fr and claire.berger@cnrs.fr

**Author Contributions**

#VO and AJ contributed equally to this work.

**Notes**

The authors have no conflicts to disclose.

ACKNOWLEDGMENT

This work is supported by a grant from the Agence Nationale de la Recherche, BONNEG project, (No ANR-19-CE24-0025). CB, AJ and AL acknowledge funding from the European Union Graphene Flagship grant agreements No. 696656 and No 785219. This work was also made possible by the French American Cultural Exchange council through a Thomas Jefferson grant.

REFERENCES

1. Backes, C. et al. Production and processing of graphene and related materials. 2d Mater 7, (2020).

2. Bonaccorso, F., Sun, Z., Hasan, T. & Ferrari, A. C. Graphene photonics and optoelectronics. Nat Photonics 4, (2010).

3. Berger, C. et al. Ultrathin epitaxial graphite: 2D electron gas properties and a route toward graphene-based nanoelectronics. Journal of Physical Chemistry B 108, 19912–19916 (2004).

4. Berger, C. et al. Electronic Confinement and Coherence in Patterned Epitaxial Graphene. https://www.science.org.

5. Emtsev, K. V. et al. Towards wafer-size graphene layers by atmospheric pressure graphitization of silicon carbide. Nat Mater 8, 203–207 (2009).

6. Forti, S. & Starke, U. Epitaxial graphene on SiC: From carrier density engineering to quasi-free standing graphene by atomic intercalation. J Phys D Appl Phys 47, (2014).

7. Baringhaus, J. et al. Exceptional ballistic transport in epitaxial graphene nanoribbons. Nature 506, (2014).

8. Dlubak, B. et al. Highly efficient spin transport in epitaxial graphene on SiC. Nat Phys 8, (2012).

9. Lin, Y. M. et al. Wafer-scale graphene integrated circuit. Science (1979) 332, (2011).

10. Tzalenchuk, A. et al. Towards a quantum resistance standard based on epitaxial graphene. Nat Nanotechnol 5, (2010).

11. Zhao, J. et al. Ultrahigh-mobility semiconducting epitaxial graphene on silicon carbide. Nature 625, 60–65 (2024).

12. Bresnehan, M. S. et al. Integration of hexagonal boron nitride with quasi-freestanding epitaxial graphene: Toward wafer-scale, high-performance devices. ACS Nano 6, 5234–5241 (2012).

13. Dean, C. R. et al. Boron nitride substrates for high-quality graphene electronics. Nat Nanotechnol 5, 722–726 (2010).

14. Fukamachi, S. et al. Large-area synthesis and transfer of multilayer hexagonal boron nitride for enhanced graphene device arrays. Nature Electronics 2023 6:2 6, 126–136 (2023).

15. Gigliotti, J. et al. Highly ordered boron nitride/epigraphene epitaxial films on silicon carbide by lateral epitaxial deposition. ACS Nano 14, 12962–12971 (2020).

16. Lipp, A., Schwetz, K. A. & Hunold, K. Hexagonal boron nitride: Fabrication, properties and applications. J Eur Ceram Soc 5, (1989).

17. Kim, K. S. et al. Non-epitaxial single-crystal 2D material growth by geometric confinement. Nature 614, 88–94 (2023).

18. Kruskopf, M. et al. Comeback of epitaxial graphene for electronics: Large-area growth of bilayer-free graphene on SiC. 2d Mater 3, (2016).

19. De Heer, W. A. et al. Large area and structured epitaxial graphene produced by confinement controlled sublimation of silicon carbide. Proc Natl Acad Sci U S A 108, 16900–16905 (2011).

20. Berger, C. et al. Epitaxial Graphene on SiC: 2D Sheets, Selective Growth, and Nanoribbons. in Growing Graphene on Semiconductors 181–204 (Jenny Stanford Publishing, 2017).

21. Ottapilakkal, V. et al. Thermal stability of thin hexagonal boron nitride grown by MOVPE on epigraphene. J Cryst Growth 603, 127030 (2023).

22. Li, X. et al. Large-area two-dimensional layered hexagonal boron nitride grown on sapphire by metalorganic vapor phase epitaxy. Cryst Growth Des 16, 3409–3415 (2016).

23. Ferrari, A. C. et al. Raman spectrum of graphene and graphene layers. Phys Rev Lett 97, (2006).

24. Hu, Y. et al. Structured epitaxial graphene: growth and properties. J Phys D Appl Phys 45, 154010 (2012).

25. Röhrl, J. et al. Raman spectra of epitaxial graphene on SiC(0001). Appl Phys Lett 92, (2008).

26. Dąbrowska, A. K. et al. Two stage epitaxial growth of wafer-size multilayer h-BN by metal-organic vapor phase epitaxy - A homoepitaxial approach. 2d Mater 8, (2020).

27. Pakuła, K. et al. Fundamental mechanisms of hBN growth by MOVPE. arXiv preprint arXiv:1906.05319 (2019).

28. Kobayashi, Y., Akasaka, T. & Makimoto, T. Hexagonal boron nitride grown by MOVPE. J Cryst Growth 310, 5048–5052 (2008).

29. Chou, H. et al. Dependence of h-BN Film Thickness as Grown on Nickel Single-Crystal Substrates of Different Orientations. ACS Appl Mater Interfaces 10, 44862–44870 (2018).

30. Lee, Y. H. et al. Growth selectivity of hexagonal-boron nitride layers on Ni with various crystal orientations. RSC Adv 2, 111–115 (2012).

31. Henry, A., Chubarov, M., Czigány, Z., Garbrecht, M. & Högberg, H. Early stages of growth and crystal structure evolution of boron nitride thin films. in Japanese Journal of Applied Physics vol. 55 (Japan Society of Applied Physics, 2016).

32. Mendelson, N. et al. Identifying carbon as the source of visible single-photon emission from hexagonal boron nitride. Nat Mater 20, (2021).

33. Ayari, T. et al. Novel Scalable Transfer Approach for Discrete III-Nitride Devices Using Wafer-Scale Patterned h-BN/Sapphire Substrate for Pick-and-Place Applications. Adv Mater Technol 4, (2019).

34. Wu, X., Li, X., Song, Z., Berger, C. & De Heer, W. A. Weak antilocalization in epitaxial graphene: Evidence for chiral electrons. Phys Rev Lett 98, (2007).

35. Beenakker, C. W. J. & van Houten, H. Quantum Transport in Semiconductor Nanostructures. Solid State Physics - Advances in Research and Applications 44, (1991).

36. Van der Pauw, L. J. A method of measuring the resistivity and Hall effect of disks of arbitrary shape. Philips Technical Review vol. 20 Preprint at (1958).

37. Novoselov, K. S. et al. Two-dimensional gas of massless Dirac fermions in graphene. Nature 438, 197–200 (2005).

38. Zhang, Y., Tan, Y. W., Stormer, H. L. & Kim, P. Experimental observation of the quantum Hall effect and Berry's phase in graphene. Nature 438, (2005).

39. Datta, S. Electronic Transport in Mesoscopic Systems. Cambridge Studies in Semiconductor Physics and Microelectronic Engineering (Cambridge University Press, 1995). doi:10.1017/CBO9780511805776.

40. Shivaraman, S., Jobst, J., Waldmann, D., Weber, H. B. & Spencer, M. G. Raman spectroscopy and electrical transport studies of free-standing epitaxial graphene: Evidence of an AB-stacked bilayer. Phys Rev B Condens Matter Mater Phys 87, (2013).

41. Riedl, C., Coletti, C., Iwasaki, T., Zakharov, A. A. & Starke, U. Quasi-free-standing epitaxial graphene on SiC obtained by hydrogen intercalation. Phys Rev Lett 103, (2009).

42. Ristein, J., Mammadov, S. & Seyller, T. Origin of doping in quasi-free-standing graphene on silicon carbide. Phys Rev Lett 108, (2012).